\documentclass[12pt]{article}

\usepackage{graphics,amssymb,epsfig,float}
\usepackage[usenames,dvips]{color}
\usepackage{graphicx}
\usepackage{epsfig}
\usepackage{rotating}
\usepackage{dcolumn}
\usepackage{bm}
\usepackage{cite}
\usepackage{amsmath}
\usepackage{setspace}

\def\lsim{\;\raise0.3ex\hbox{$<$\kern-0.75em\raise-1.1ex\hbox{$\sim$}}\;}
\def\gsim{\;\raise0.3ex\hbox{$>$\kern-0.75em\raise-1.1ex\hbox{$\sim$}}\;}
\def\beq{\begin{equation}}   \def\eeq{\end{equation}}
\def\ba{\begin{array}}       \def\ea{\end{array}}
\def\bea{\begin{eqnarray}}   \def\eea{\end{eqnarray}}

\def\nl{\newline}

\begin{document}

\begin{titlepage}
\begin{flushright}
LPT Orsay 13-69
\end{flushright}

\begin{center}
\vspace{1cm}
{\Large\bf Testing the higgsino-singlino sector of the NMSSM\\[3mm]
with trileptons at the LHC}
\vspace{2cm}

{\bf{Ulrich Ellwanger}}
\vspace{1cm}\\
Laboratoire de Physique Th\'eorique, UMR 8627,\\ CNRS and
Universit\'e de Paris--Sud,
F-91405 Orsay, France

\end{center}

\vspace{1cm}

\begin{abstract}
We propose a simplified light higgsino-singlino scenario in the NMSSM,
in which the masses of the chargino and the lightest neutralino
determine the masses and couplings of all 3 lightest neutralinos. This
scenario is complementary to the simplified wino-like
chargino/neutralino scenario used conventionally for the interpretation
of results from trilepton searches, and motivated by lower bounds on the
gluino mass in the case of GUT relations between the wino and gluino
masses. We present all masses and mixing angles necessary for the
determination of production cross sections of the chargino and the
3~neutralinos in the form of Tables in the $M_{\chi^0_1}-M_{\chi^\pm_1}$
plane, assuming Higgs mass motivated values for $\tan\beta=2$ and
$\lambda=0.6$. We show that this scenario leads to considerable signal
rates, and present constraints in this plane from recent searches for
trileptons at the LHC.

\end{abstract}

\end{titlepage}

\newpage
\section{Introduction}
One of the main tasks of the LHC is the search for supersymmetric (SUSY)
particles. These searches have been without success so far, and have
lead to lower bounds in the TeV range on the masses of gluinos and
squarks of the first two generations.

The trilepton channel $pp \to W^{(*)} \to \chi^\pm + \chi^0_i$ ($i > 1$)
with $\chi^\pm \to W^{(*)} + \chi^0_1$, $\chi^0_i \to Z^{(*)} +
\chi^0_1$ and leptonic decays of both $W^{(*)}$ and $Z^{(*)}$ are
considered as ``gold-plated'' for searches for charginos and neutralinos
at hadron colliders \cite{Nath:1987sw,Arnowitt:1987pm,Barbieri:1991vk,
Baer:1992dc,Lopez:1992ss,Baer:1993tr,Baer:1994nr,Baer:1995va,
Baer:2012wg}. Corresponding searches at the LHC using leptonic final
states, in particular trileptons plus missing transverse energy, have
been performed by ATLAS and CMS \cite{CMS-JHEP06,ATLAS-CONF-2012-154,
Aad:2012jja,Aad:2012hba,Chatrchyan:2012pka,CMS-PAS-SUS-12-022}; see
\cite{ATLAS:2013rla,CMS-PAS-SUS-13-006} for recent results from
trilepton searches at $\sqrt{s}=8$\;TeV. There, the absence of
significant excesses of events is interpreted in simplified models,
motivated by the SUSY particle content of the Minimal SUSY extension of
the Standard Model (MSSM) and assuming simple branching ratios and
relations among masses. Interpretations of the present constraints from
the trilepton channel in realistic versions of the MSSM have been
performed in \cite{AbdusSalam:2012ir,Bharucha:2013epa,Kowalska:2013ica,
Belanger:2013pna}
(see \cite{Frank:2012ne,Cheng:2013fma,Alloul:2013fra} for the r\^ole of
trilepton final states in some beyond-the-MSSM models). 

A particularly challenging MSSM scenario would be the case of light and
nearly degenerate higgsinos \cite{Baer:2011ec,Bobrovskyi:2011jj}. Light
higgsinos correspond to a relatively small ($\lesssim 300$~GeV) SUSY Higgs
mass parameter $\mu$, which is favored by fine-tuning arguments: $\mu^2$
appears always as a positive mass$^2$ parameter in the Higgs potential,
and must be compensated by negative soft SUSY breaking Higgs mass terms
for electroweak symmetry breaking to be possible. This makes large
values of $\mu$ unnatural, since then the required cancellation must be
relatively very precise. On the other hand, a dominantly higgsino-like
(neutralino) LSP is a difficult candidate for dark matter: its
annihilation cross section in the early universe is typically too large
so that, assuming a standard cosmological evolution, its relic density
today is too small to comply with the WMAP and Planck result $\Omega h^2
\sim 0.1187$. Moreover, its direct detection (nucleon) cross section is
also too large \cite{Perelstein:2011tg,Belanger:2013pna} in view of the
latest XENON100 results \cite{Aprile:2012nq} unless the higgsino
component is close to 100\%.

In the Next-to-Minimal Supersymmetric Standard Model (NMSSM)
\cite{Ellwanger:2009dp}, a SUSY Higgs mass parameter $\mu_{\text{eff}}$
is generated dynamically through the vacuum expactetion value (vev) of a
gauge singlet superfield $S$, $\mu_{\text{eff}} = \lambda \left<
S\right>$, where $\lambda$ denotes the coupling of $S$ to the MSSM-like
Higgs superfields. Again, small values of $\mu_{\text{eff}}$ are favored
by low fine-tuning \cite{Ellwanger:2011mu,Perelstein:2012qg}. However,
the problems associated with a higgsino-like LSP in the MSSM do not
persist in the NMSSM due to the presence of the additional neutralino
(singlino) from the superfield $S$, which can well be lighter than the
higgsinos. This allows for a dominantly singlino-like LSP, whose
annihilation cross section via Higgs bosons in the s-channel (and/or
coannihilation or some higgsino component) can lead to the desired relic
density (see \cite{Ellwanger:2009dp} and refs. therein), and a
sufficiently small direct detection cross section 
\cite{Cerdeno:2004xw,Cerdeno:2007sn,Barger:2007nv,Belanger:2008nt}
compatible with present XENON100 bounds.

Hence a scenario with light higgsinos and a light singlino, but heavy
electroweak gauginos, is viable in the NMSSM. Assuming GUT relations
among the electroweak gauginos and the gluino, heavy electroweak
gauginos would be an obvious consequence of lower bounds well above
1~TeV on the gluino mass. Since higgsino-singlino mixing would lift the
degeneracy among the neutral and charged higgsinos and in the presence
of a light singlino-like LSP, searches for charginos and neutralinos in
the trilepton channel would be more promising than in the light higgsino
scenario within the MSSM  \cite{Baer:2011ec}.

We emphasize that such a scenario would be complementary to the
simplified models used by ATLAS and CMS for the interpretation of the
trilepton results up to now: In these models, the lightest chargino and
the second lightest neutralino are assumed to be wino-like and
degenerate, the lightest neutralino is assumed to be bino-like, and the
higgsinos are assumed to be heavy and decoupled. The obvious reason for
this choice is that now signal rates depend on only two mass parameters
(assuming 100\% branching ratios into $W/Z + \chi^0_1$).

In the light higgsino-singlino scenario of the NMSSM (containing one
chargino $\chi^\pm_1$ and 3 neutralinos $\chi^0_i$, $i=1,2,3$) the
masses and neutralino mixing angles depend on the higgsino mass
parameter $\mu_{\text{eff}}$, the singlino mass, the ratio $\tan\beta$
of Higgs vevs and the coupling $\lambda$ (see the next Section). The
value of $\sim 125$~GeV for the SM-like Higgs mass can be obtained
naturally in the NMSSM provided $\lambda$ is large ($\approx 0.6$) and
$\tan\beta$ is relatively small ($\approx 2$) \cite
{Hall:2011aa,Ellwanger:2011aa,Arvanitaki:2011ck,King:2012is,Kang:2012sy,
Cao:2012fz,Ellwanger:2012ke,Jeong:2012ma,Randall:2012dm,Benbrik:2012rm,
Kyae:2012rv,Cao:2012yn,Agashe:2012zq,Belanger:2012tt,Heng:2012at,
Choi:2012he,King:2012tr,Gherghetta:2012gb,Cheng:2013fma,
Barbieri:2013hxa,Badziak:2013bda,Cheng:2013hna,Hardy:2013ywa,Beskidt:2013gia}.
Fixing $\tan\beta=2$ and $\lambda=0.6$, also the light higgsino-singlino
scenario of the NMSSM depends on two mass parameters only. These can be
chosen as the physical chargino mass $M_{\chi^\pm_1}$ ($\sim
\mu_{\text{eff}}$) and the LSP mass $M_{\chi^0_1}$, which have a more
direct physical meaning than the higgsino and singlino mass parameters
in the Lagrangian.

Assuming heavy sleptons, the chargino $\chi^\pm_1$ decays with a
branching ratio (BR) of 100\% into $W^{(*)} + \chi^0_1$. The neutralinos
$\chi^0_{2,3}$ will typically decay with a BR of 100\% into  $Z^{(*)} +
\chi^0_1$. In some cases, decays $\chi^0_{2,3} \to \chi^0_1 +$ a CP-odd
or CP-even Higgs boson are possible. These depend on the masses and
couplings of the Higgs bosons. Subsequently we define a simplified light
higgsino-singlino scenario within which such neutralino-to-Higgs decays
are absent. Then, given the above values of $\tan\beta$ and $\lambda$,
$M_{\chi^\pm_1}$ and $M_{\chi^0_1}$ determine completely the signal
rates for chargino + neutralino production into trilepton final states.

It would be very helpful if the ATLAS and CMS collaborations would
interpret their results from trilepton searches within such a light
higgsino-singlino scenario, as it would allow to test a well motivated
region in the parameter space of the NMSSM. It also allows to generalize
in a well defined manner the (somewhat unrealistic) assumption
$M_{\chi^\pm_1}=M_{\chi^0_2}$ within the present simplified models, and
to study the impact of the lift of this degeneracy.

In the present paper we present the necessary parameters for the
determination of the $pp \to \chi^\pm_1+\chi^0_{2,3}$ production rates
in the light higgsino-singlino scenario in the form of Tables in the
$M_{\chi^\pm_1}-M_{\chi^0_1}$ plane: The masses $M_{\chi^0_2}$ and
$M_{\chi^0_3}$ of the two additional neutralinos, and the mixing angles
of $\chi^0_2$ and $\chi^0_3$ (relevant for the
$W-\chi^\pm_1-\chi^0_{2,3}$ couplings). As a (preliminary) application
we simulated the trilepton event rates in the various signal channels
defined in \cite{ATLAS:2013rla}, and deduced bounds in the
$M_{\chi^\pm_1}-M_{\chi^0_1}$ plane from the upper limits on the
corresponding channel-specific signal cross sections based on
21~fb$^{-1}$ integrated luminosity at $\sqrt{s}=8$~TeV given there. The
Tables should allow the ATLAS and CMS collaborations to perform more
precise simulations (in particular of the detector responses) and
comparisons with data
themselves, notably with future data at larger $\sqrt{s}$ and/or
different signal channels.

In the next Section we define the simplified light higgsino-singlino
scenario in the NMSSM, and comment on the Tables given in the Appendix.
In Section~3 we describe the simulations of the trilepton final states
and show the resulting bounds in the $M_{\chi^\pm_1}-M_{\chi^0_1}$
plane. Section~4 is devoted to conclusions and an outlook.

\section{The Light Higgsino-Singlino Scenario in the NMSSM}

The NMSSM differs from the MSSM due to the presence of the gauge singlet
superfield~$\hat S$. In the simplest realisation of the NMSSM, the SUSY
$\mu \hat H_u \hat H_d$ Higgs mass term in the MSSM superpotential
$W_\text{MSSM}$  is replaced by the coupling $\lambda$ of $\hat S$ to
$\hat H_u$ and $\hat H_d$, and a self-coupling $\kappa \hat S^3$.  
Here, $\hat H_u$ couples to up-type quark superfields, and $\hat H_d$ to
down-type quark and lepton superfields. In this version the
superpotential $W_\text{NMSSM}$ is scale invariant, and given by:
\beq\label{eq:1}
W_\text{NMSSM} = \lambda \hat S \hat H_u\cdot \hat H_d +
\frac{\kappa}{3}  \hat S^3 +\dots
\eeq
where the dots denote the Yukawa couplings of $\hat H_u$ and $\hat H_d$
to the quarks and leptons as in the MSSM. Once the scalar component of
$\hat S$ develops a vev $s$, the first term in $W_\text{NMSSM}$
generates an effective $\mu$-term with
\beq\label{eq:2}
\mu_\mathrm{eff}=\lambda\, s\; .
\eeq
Amongst others, $\mu_\mathrm{eff}$ generates a Dirac mass term
$\mu_\mathrm{eff} \psi_u \psi_d+ \mathrm{h.c.}$ for the SU(2)-doublet
higgsinos $\psi_u$ and $\psi_d$. The vevs $v_u$ and $v_d$ of $\hat H_u$
and $\hat H_d$ generate mixing terms between the singlino $\psi_S$ and
$\psi_d$, $\psi_u$, respectively. The second term in the superpotential
$W_\text{NMSSM}$ \eqref{eq:1} generates a Majorana mass term $M_S$ for
$\psi_S$ with
\beq\label{eq:3}
M_S = 2\, \kappa\, s \equiv \frac{2 \kappa}{\lambda}\mu_\mathrm{eff} 
\; .
\eeq
The soft SUSY breaking terms include, amongst others,
mass terms for the gauginos $\tilde{B}$ (bino), $\tilde{W}^a$
(winos) and $\widetilde{G}^a$ (gluinos):
 \beq\label{eq:4}
-{\cal L}_\mathrm{1/2}= \frac{1}{2} \bigg[ 
 M_1 \widetilde{B}  \widetilde{B}
\!+\!M_2 \sum_{a=1}^3 \widetilde{W}^a \widetilde{W}_a 
\!+\!M_3 \sum_{a=1}^8 \widetilde{G}^a \widetilde{G}_a   
\bigg]+ \mathrm{h.c.}\;.
\eeq
Altogether the symmetric $5 \times 5$ mass matrix ${\cal M}_0$
in the neutralino sector in the basis $\psi^0 = (-i\widetilde{B} ,
-i\widetilde{W}^3, \psi_d^0, \psi_u^0, \psi_S)$
is given by \cite{Ellwanger:2009dp}
\beq\label{eq:5}
{\cal M}_0 =
\left( \ba{ccccc}
M_1 & 0 & -\frac{g_1 v_d}{\sqrt{2}} & \frac{g_1 v_u}{\sqrt{2}} & 0 \\
& M_2 & \frac{g_2 v_d}{\sqrt{2}} & -\frac{g_2 v_u}{\sqrt{2}} & 0 \\
& & 0 & -\mu_\mathrm{eff} & -\lambda v_u \\
& & & 0 & -\lambda v_d \\
& & & & M_S
\ea \right)
\eeq
leading to mass terms in the Lagrangian of the form
\beq\label{eq:6}
{\cal L} = - \frac{1}{2} (\psi^0)^T {\cal M}_0 (\psi^0) + \mathrm{h.c.}\; .
\eeq

In the limit of heavy decoupled winos and bino considered here, $M_{1,2}
\gg  \left(\mu_\mathrm{eff},\ 2|\kappa s|\right)$ (we assume
$\mu_\mathrm{eff} > 0$), the light higgsino-singlino sector is described
by the lower $3 \times 3$ sub-matrix ${\cal M}_0^{(3)}$ of \eqref{eq:5}.
Besides $\mu_\mathrm{eff}$ and $M_S$, the elements of ${\cal M}_0^{(3)}$
depend on $\lambda$ and $\tan\beta = \frac{v_u}{v_d}$ via $v_u =
\sin\beta \sqrt{v_u^2+v_d^2} =\sin\beta (174\ \text{GeV})$, $v_d =
\cos\beta \sqrt{v_u^2+v_d^2}$. As stated in the introduction, we define
subsequently the simplified light higgsino-singlino scenario by
$\lambda=0.6$, $\tan\beta = 2$.

${\cal M}_0^{(3)}$ is diagonalized by an orthogonal real matrix
$N_{ij}$, $i=1,2,3$, $j=3,4,5$ such that the physical masses
$M_{\chi^0_i}$ ordered in $|M_{\chi^0_i}|$ are real, but not necessarily
positive. Denoting the 3 eigenstates by $\chi^0_i$, we have
\beq\label{eq:7}
\chi^0_i = N_{ij} \psi^0_j
\eeq
where $\psi^0_3 = \psi_d^0$, $\psi^0_4 = \psi_u^0$, $\psi^0_5 = 
\psi_S$.

The Dirac mass $M_{\chi^\pm_1}$ of the higgsino-like charginos
$\chi^\pm_1$ is simply $M_{\chi^\pm_1} = \mu_\mathrm{eff}$.

The calculations of the masses and mixing angles as function of the
parameters $\lambda$, $\kappa$ (which determines $M_S$),
$\mu_\mathrm{eff}$ and $\tan\beta$ are performed with the help of the
public code
{\textsf{NMSSMTools}}~\cite{Ellwanger:2004xm,Ellwanger:2005dv}. Due to
the radiative corrections to the pole masses, $M_{\chi^\pm_1}$ differs
slightly from $\mu_\mathrm{eff}$. For the soft SUSY breaking squark and
slepton masses we choose 2~TeV, and $M_1=1$~TeV, $M_2=2$~TeV,
$M_3=6$~TeV.  (These SUSY breaking parameters determine implicitely the
SUSY breaking scale which has a mild impact on the radiative
corrections.)

The soft SUSY breaking parameters $A_\lambda$ and $A_\kappa$ 
\cite{Ellwanger:2009dp} have no impact on the neutralino/\-char\-gino
sector. They can be chosen such that the SM-like Higgs mass is near
125~GeV; in any case they must be chosen such that all physical Higgs
masses$^2$ are positive.

Subsequently we consider $M_{\chi^\pm_1}$ in the range
$M_{\chi^\pm_1}=100\dots 400$~GeV in steps of 20~GeV. The corresponding
values of $\mu_\mathrm{eff}$ are tabulated in Table~\ref{t:mu} in the
Appendix.

For $M_{\chi^0_1}$ we consider the range $M_{\chi^0_1}=0\dots 100$~GeV
in steps of 10~GeV. For fixed $\lambda$, $\tan\beta$ and for each given
$\mu_\mathrm{eff}$, the desired values for $M_{\chi^0_1}$ can be
obtained by suitable values of $\kappa$. Actually, low values of
$M_{\chi^0_1}$  require negative values for $\kappa$ (we recall that we
assume $\mu_\mathrm{eff}>0$). Along a strip around $\kappa \sim 0$,
negative and positive values of $\kappa$ (of different absolute values)
can lead to the same values of $M_{\chi^0_1}$. We chose the convention
$\kappa>0$ whenever possible, which lifts this ambiguity. The
corresponding values for $\kappa$ are tabulated in Table~\ref{t:k} in
the Appendix. Note that $M_{\chi^0_1}=M_{\chi^\pm_1}$ (= 100~GeV) is not
possible for $M_S \leq \mu_\mathrm{eff}$ as considered here, since even
for $M_S = \mu_\mathrm{eff}$ mixing in the neutralino sector will always
imply $M_{\chi^0_1}<M_{\chi^\pm_1}$ (by at least 3~GeV).

Now the masses and mixing angles of $\chi^0_2$ and $\chi^0_3$ are
uniquely determined. In the Tables~\ref{t:m2} and \ref{t:m3} in the
Appendix we list $M_{\chi^0_2}$ and $M_{\chi^0_3}$ in the considered
range of $M_{\chi^\pm_1}$ and $M_{\chi^0_1}$. We see that $M_{\chi^0_2}$
is close to  $M_{\chi^\pm_1}\sim \mu_\mathrm{eff}$, whereas
$M_{\chi^0_3}$ is significantly larger due to mixing. (For $M_{\chi^0_1}
\gsim M_{\chi^\pm_1} - 20$~GeV in the lower left-hand corner, the mixing
angles differ considerably from the other part of the plane.)

The mixing angles $N_{i,j}$ with $i=2,3$, $j=3,4$ appear in the Feynman
rule for the vertex in $W^+ \to \chi^+_1 + \chi^0_{2,3}$ corresponding
to an incoming $W^+_\mu$, an outgoing  higgsino-like chargino $\chi^+_1$
and outgoing neutralinos $\chi^0_{2,3}$; the projectors $P_L$, $P_R$ act
on the neutralino 4-spinors:
\beq\label{eq:8}
i g_2 \gamma_\mu (N_{i,3} P_R - N_{i,4} P_L)/\sqrt{2}\; .
\eeq
Note that the last factor $1/\sqrt{2}$ is absent in the coupling of $W$
to wino-like charginos and neutralinos (not shown here); as a
consequence the production rates of higgsino-like charginos and
neutralinos considered here are smaller than those of wino-like
charginos and neutralinos assumed in the simplified models used by ATLAS
and CMS. The mixing angles $N_{2,3}$, $N_{2,4}$, $N_{3,3}$ and $N_{3,4}$
are tabulated in the Tables~\ref{t:n23}, \ref{t:n24}, \ref{t:n33} and
\ref{t:n34}, respectively, in the Appendix. Since the signs of physical
spinors can be flipped without affecting the cross sections, we used
this freedom to simplify the tables in the lower left-hand corner where
frequent sign changes appear as consequence of the numerical
diagonalisation routine.

Given the decoupled winos and bino, the (absolute values of) the
singlino components of $\chi^0_{i}$, $i=2,3$, are given by $|N_{i,5}| =
\sqrt{1-N_{i,3}^2 - N_{i,4}^2}$, and are typically quite small with the
exception of $N_{3,5}$ for $M_{\chi^0_1} \lsim M_{\chi^\pm_1}$ (in the
lower left-hand corner of the Tables). Likewise, the (absolute values
of) the higgsino components $N_{1,i}$ ($i=3,4$) of $\chi^0_{1}$ are
given by $|N_{1,i}| = \sqrt{1-N_{2,i}^2 - N_{3,i}^2}$, from which the
singlino component $N_{1,5}$ of $\chi^0_{1}$ (typically dominant) can be
deduced as before.

Assuming BRs of 100\% for the decays $\chi^0_2 \to  Z^{(*)} + \chi^0_1$,
$\chi^0_3 \to  Z^{(*)} + \chi^0_1$ and $\chi^\pm_1 \to  W^{(*)} +
\chi^0_1$, the masses and the $W-\chi^\pm_1-\chi^0_{2,3}$ vertices
suffice to determine the production cross sections and signal rates into
trilepton channels in this light higgsino-singlino scenario. Hence this
scenario is now well defined (on the points given in the Tables), and
can be used to interpret results from searches for trileptons and
missing energy. If desired, the parameters of additional points in the 
$M_{\chi^0_1}-M_{\chi^\pm_1}$ plane can be provided.

\section{Present LHC Constraints on the Light Higgsino-\\ Singlino
Scenario}

Recent results from searches for trileptons and missing energy at the
LHC at $\sqrt{s}=8$~TeV and $\sim 20$~fb$^{-1}$ of integrated luminosity
have been published in \cite{ATLAS:2013rla,CMS-PAS-SUS-13-006}.
Subsequently we use the upper bounds (at the 95\% confidence level) on
event rates in the specific search channels given by ATLAS in the
Table~1 in \cite{ATLAS:2013rla}. For the simulation we proceed as
follows:

The cross sections for $pp \to \chi^\pm_1 + \chi^0_2$ and  $pp \to
\chi^\pm_1 + \chi^0_3$ production are obtained from Prospino at
next-to-leading order (NLO) \cite{Beenakker:1996ch,
Beenakker:1996ed,Beenakker:1999xh}. The matrix elements are generated by
MadGraph/MadEvent~5 \cite{Alwall:2011uj}, which includes Pythia~6.4
\cite{Sjostrand:2006za}. The output is given to the fast detector
simulation DELPHES \cite{Ovyn:2009tx,deFavereau:2013fsa}. We have
verified that the cut flows given in Table~5 in \cite{ATLAS:2013rla} are
reproduced reasonably well (within $10-30\%$).

The strongest constraints on the light higgsino-singlino scenario
originate from the search channels SRnoZc and SRZc in
\cite{ATLAS:2013rla}, which are defined as follows: $m_\text{SFOS}$
denotes the invariant mass of a same-flavour opposite-sign (SFOS) lepton
pair (closest to the $Z$-boson mass). For SRnoZc one requires
$m_\text{SFOS} < 81.2$~GeV or $m_\text{SFOS} > 101.2$~GeV, whereas for
SRZc one requires $81.2~\text{GeV} < m_\text{SFOS} < 101.2$~GeV. For
$p_T^l$ of the third leading lepton one requires $p_T^l > 30$~GeV for
SRnoZc, and $p_T^l > 10$~GeV for SRZc. For the missing transverse energy
$E_T^\text{miss}$ one requires $E_T^\text{miss} > 75$~GeV for SRnoZc,
and $E_T^\text{miss} > 120$~GeV for SRZc. The transverse mass $m_T$ is
defined by $m_T=\sqrt{2\cdot E_T^\text{miss}\cdot p_T^l\cdot(1-\cos
\Delta\Phi_{l,E_T^\text{miss}})}$, and required to be above 110~GeV in
all cases. The upper bounds on the number of signal events are 6.8 for
SRnoZc, and 6.5 for SRZc (for more details, see \cite{ATLAS:2013rla}).

\begin{figure}
\begin{center}\includegraphics[scale=1.15,clip=true]{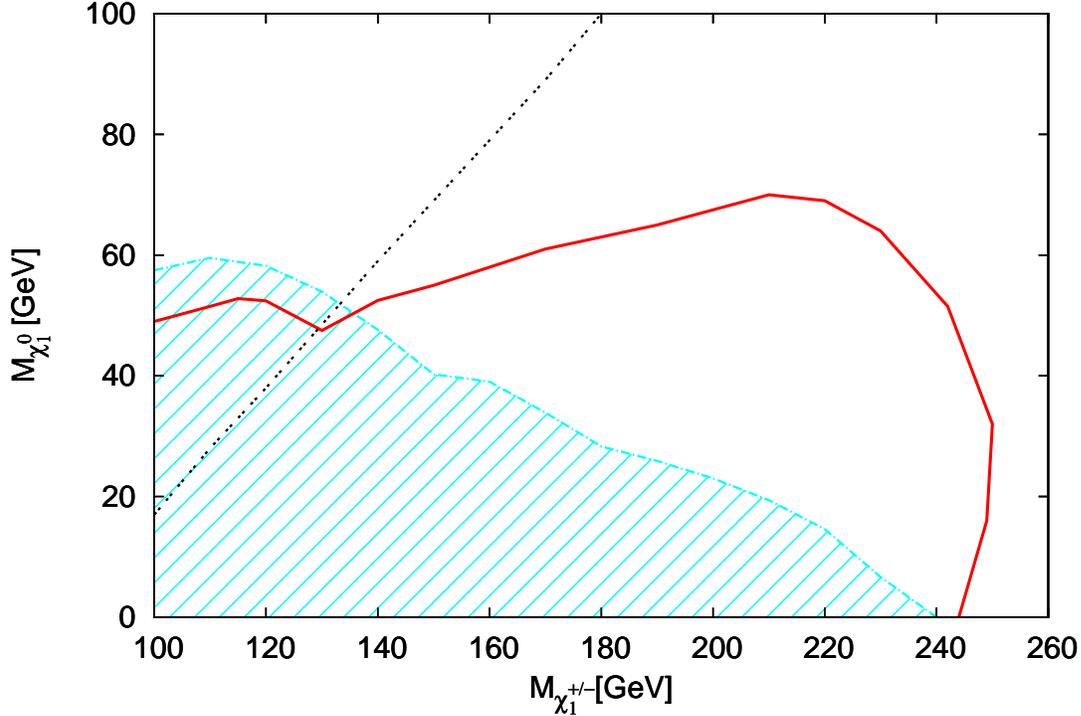}
\end{center}
\caption{
Red curve: Boundary of the excluded region in the
$M_{\chi^0_1}-M_{\chi^\pm_1}$ plane from searches for trileptons by
ATLAS \cite{ATLAS:2013rla}.  The black dashed line indicates
$M_{\chi^0_2}-M_{\chi^0_1}=M_Z$. Above the black dashed line, $\chi^0_2$
undergoes 3-body decays $\chi^0_2 \to \chi^0_1+Z^* \to ...$. The blue
hashed region is excluded by searches for $\chi^0_2+\chi^0_1$ production
at LEP \cite{Abdallah:2003xe,Abbiendi:2003sc}.}
\label{fig:1}
\end{figure}

The resulting constraints in the $M_{\chi^0_1}-M_{\chi^\pm_1}$ plane are
shown as a red curve in Fig.~\ref{fig:1}. The black dashed line in
Fig.~\ref{fig:1} indicates where $M_{\chi^0_2}-M_{\chi^0_1}=M_Z$. Above
the black dashed line, $\chi^0_2$ (with a significantly larger
production cross section than $\chi^0_3$) undergoes 3-body decays
$\chi^0_2 \to \chi^0_1+Z^* \to ...$, and the search channel SRnoZc is
most relevant, whereas below the black dashed line $\chi^0_2$ undergoes
2-body decays $\chi^0_2 \to \chi^0_1+Z \to ...$, and the search channel
SRZc is relevant.

We note that the bounds in the $M_{\chi^0_1}-M_{\chi^\pm_1}$ plane are
significantly weaker than the bounds in Fig.~8(b) in
\cite{ATLAS:2013rla}. The reason herefore is the significantly lower
production cross section for higgsino-like charginos and neutralinos
compared to wino-like charginos and neutralinos assumed in the
simplified model in \cite{ATLAS:2013rla}. The fact that here we include
in addition the production of $\chi^0_3$ does not compensate for the
smaller couplings to $W^{*}$ (also due to mixings with the singlino),
moreover $M_{\chi^0_3}$ is substantially larger than $M_{\chi^0_2}$.

Searches for $e^+ e^- \to Z^{(*)} \to \chi^0_2 \chi^0_1$ had already
been preformed at LEP, notably by the DELPHI \cite{Abdallah:2003xe} and
OPAL \cite{Abbiendi:2003sc} collaborations. Since the $Z- \chi^0_2
-\chi^0_1$ vertex is equally well defined in the present scenario,
corresponding constraints (as implemented in
{\textsf{NMSSMTools}}~\cite{Ellwanger:2004xm,Ellwanger:2005dv}) can be
applied and lead to the excluded blue hashed region in Fig.~\ref{fig:1}.
Actually, inside the blue hashed region (for $M_{\chi^0_1}< M_Z/2$ and
low $M_{\chi^\pm_1}$ where the higgsino components of $ \chi^0_1$ are
not negligibly small), some points are also excluded by a too large
contribution to the invisible $Z$~width from $Z \to \chi^0_1\chi^0_1$.

Hence, even in this somewhat delicate light higgsino-singlino scenario,
already the present LHC constraints are significantly stronger than the
bounds from LEP.

\section{Conclusions and outlook}

The main purpose of the present paper is the presentation of a
simplified, but well-motivated light higgsino-singlino scenario in the
NMSSM. It is complementary to the simplified wino-like
chargino/neutralino scenario used conventionally by ATLAS and CMS for
the interpretation of results from trilepton searches, but leads equally
to considerable signal rates. (Assuming GUT-like relations among the
gaugino mass parameters, the wino mass term is bounded from below by
about $1/3$ of the lower limit on the gluino mass, disfavouring light
winos.) Hence present and future searches for chargino/neutralino
production at the LHC merit to be interpreted in this scenario as well.

We have presented all necessary masses and mixing angles for the
determination of production cross sections of the chargino and the
3~neutralinos in the form of Tables in the $M_{\chi^0_1}-M_{\chi^\pm_1}$
plane, assuming Higgs mass motivated values (in the NMSSM) for
$\tan\beta=2$ and $\lambda=0.6$; however, as long as $\lambda$ remains
sizeable and $\tan\beta$ small, the masses and mixing angles depend
little on these specific values. Masses and mixing angles for additional
points in this plane can be provided by the author, or be obtained with
the help of the public code 
{\textsf{NMSSMTools}}~\cite{Ellwanger:2004xm,Ellwanger:2005dv}.

The present constraints in the $M_{\chi^0_1}-M_{\chi^\pm_1}$
plane from the ATLAS search for trileptons \cite{ATLAS:2013rla}
(assuming heavy sleptons) have been obtained using the simplified
detector simulation DELPHES \cite{Ovyn:2009tx,deFavereau:2013fsa}; this
simulation should actually be redone by the LHC collaborations
themselves. In any case it shows that the signal rates in this scenario
are measurably large, and that relevant constraints are obtained in the
absence of signals.

In addition to the trilepton final state, the simplified
higgsino-singlino scenario can also be tested in searches for four
leptons plus $E_T^{\text{miss}}$ \cite{CMS-PAS-SUS-13-006,
ATLAS-CONF-2013-036}. The relevant processes would be $pp \to Z^* \to
\chi_i^0\chi_j^0$ with $i,j=2,3$, $\chi_{i,j}^0 \to \chi_1^0+Z^{(*)}$
and leptonic decays of $Z^{(*)}$ (or, alternatively, hadronic decays of
one of the $Z^{(*)}$ \cite{CMS-PAS-SUS-13-006,
ATLAS-CONF-2012-152}). Again, the masses and mixing angles
given here allow for the determination of the production cross sections.
These are dominated by $\chi_2^0 + \chi_3^0$ production with, however, a
cross section of only $\sim 20-25\%$ of the one for trileptons. At
present, the results of these searches are interpreted in SUSY models
with gauge mediation where $\chi_1^0$ is replaced by a practically
massless gravitino. Hence, re-analysises of these (and future) results
are desirable in the context of the present scenario.

In how far does this simplified light higgsino-singlino scenario differ
from more general regions in the parameter space of the NMSSM? First, if
the wino mass parameter $M_2$ is not as large as assumed here (2~TeV),
the lightest chargino and neutralinos can have sizeable wino components
which lead typically to an increase of the production cross sections.
Hence the present scenario is conservative in this respect. On the other
hand, the neutralinos $\chi^0_{2,3}$ will not necessarily have branching
ratios of 100\% into $Z^{(*)} + \chi^0_1$, notably if decays into some
of the lighter of the 5~CP-even or CP-odd neutral Higgs states of the
NMSSM are frequent. Clearly such decays would reduce the signal rate in
the trilepton channel considered here -- even if $\chi^0_{2,3}$ decay
into a light pseudoscalar $A_1$ with $M_{A_1} < 2 M_b$ and $A_1$ decays
dominantly into $A_1 \to \tau^+\tau^-$
\cite{Cheung:2008rh,Cerdeno:2013qta}. Decays of $\chi^0_{2,3}$ into
heavier Higgs states (decaying into $b\bar{b}$), however, start also to
be constrained by corresponding searches \cite{CMS-PAS-SUS-13-017}.
Implications of these latter results on more general regions in the
parameter space of the NMSSM remain to be worked out.

\section*{Acknowledgements}

It is a pleasure to thank A.~Teixeira for discussions.
We acknowledge support from the French
ANR~LFV-CPV-LHC, the European Union FP7 ITN INVISIBLES
(Marie Curie Actions,~PITN-GA-2011-289442) and the ERC advanced grant
Higgs@LHC.


\section*{Appendix} 

Below we list in the form of Tables in the $M_{\chi^0_1}-M_{\chi^\pm_1}$
plane various parameters relevant for the simplified light
higgsino-singlino scenario in the NMSSM. First, in~Table \ref{t:mu} we
show the values of $\mu_\mathrm{eff}$ leading to specific values for
$M_{\chi^\pm_1}$ (which do not depend on $M_{\chi^0_1}$). In
Table~\ref{t:k} we give the values of $\kappa$ which lead to the
corresponding values of $M_{\chi^0_1}$. In the Tables~\ref{t:m2} and
\ref{t:m3} we list $M_{\chi^0_2}$ and $M_{\chi^0_3}$, and  in the
Tables~\ref{t:n23}, \ref{t:n24}, \ref{t:n33} and \ref{t:n34} we list
$N_{2,3}$, $N_{2,4}$, $N_{3,3}$ and $N_{3,4}$, respectively.

\begin{table}[!h]
\center
\begin{tabular}{|c|c|}
\hline
$M_{\chi^\pm_1}$ &{\fontsize{10}{1} \selectfont
100.0 120.0 140.0 160.0 180.0 200.0 220.0 240.0 260.0 280.0 300.0 
 320.0 340.0  360.0 380.0 400.0 }\\
\hline 
$\mu_\mathrm{eff}$ &{\fontsize{10}{1} \selectfont
\ 99.3 118.8 138.4 157.9 177.5 197.2 216.8 236.5 256.1 275.8 295.6 315.3 335.0 354.8 374.6 394.3 }\\
\hline
\end{tabular}
\caption{Values for $\mu_\mathrm{eff}$ leading to the specific values of
$M_{\chi^\pm_1}$ used below.}
\label{t:mu}
\end{table}

\begin{table}[!h]
\center
\begin{tabular}{|c|c|}
\hline
$M_{\chi^\pm_1}$: &{\fontsize{10}{1} \selectfont
100.0 120.0 140.0 160.0 180.0 200.0 220.0 240.0 260.0 280.0 300.0 
 320.0 340.0  360.0 380.0 400.0 }\\
\hline \hline
$M_{\chi^0_1}$ & $\kappa$ \\
\hline
  0 & {\fontsize{10}{1} \selectfont
-.242\, -.170\, -.126\, -.097\, -.077\, -.063\, -.052\, -.044\, -.037\, -.032\, -.028\, -.025\, -.022\, -.020\, -.018\, -.016 }\\
 10 & {\fontsize{10}{1} \selectfont
-.185\, -.129\, -.095\, -.072\, -.056\, -.044\, -.036\, -.029\, -.024\, -.020\, -.017\, -.015\, -.013\, -.011\, -.009\, -.008 }\\
 20 & {\fontsize{10}{1} \selectfont
-.130\, -.090\, -.064\, -.047\, -.035\, -.026\, -.020\, -.015\, -.011\, -.008\, -.006\, -.004\, -.003\, -.002\, -.001\, -.000 }\\
 30 & {\fontsize{10}{1} \selectfont
-.077\, -.051\, -.034\, -.022\, -.014\, -.008\, -.004\, -.089\, 0.002 0.003 0.005 0.006 0.006 0.007 0.007 0.008 }\\
 40 & {\fontsize{10}{1} \selectfont
-.025\, -.013\, -.004\, 0.003 0.007 0.010 0.012 0.014 0.015 0.015 0.016 0.016 0.016 0.016 0.016 0.016 }\\
 50 & {\fontsize{10}{1} \selectfont
0.029 0.025 0.026 0.027 0.028 0.028 0.028 0.028 0.027 0.027 0.026 0.026 0.025 0.025 0.024 0.023 }\\
 60 & {\fontsize{10}{1} \selectfont
0.088 0.064 0.055 0.051 0.048 0.046 0.044 0.042 0.040 0.039 0.037 0.036 0.035 0.033 0.032 0.031 }\\
 70 & {\fontsize{10}{1} \selectfont
0.159 0.105 0.086 0.075 0.068 0.063 0.059 0.056 0.053 0.050 0.048 0.046 0.044 0.042 0.041 0.039 }\\
 80 & {\fontsize{10}{1} \selectfont
0.264 0.151 0.117 0.100 0.089 0.081 0.075 0.070 0.066 0.062 0.059 0.056 0.053 0.051 0.049 0.047 }\\
 90 & {\fontsize{10}{1} \selectfont
0.539 0.207 0.150 0.125 0.110 0.099 0.091 0.084 0.079 0.074 0.070 0.066 0.063 0.060 0.057 0.055 }\\
100 & {\fontsize{10}{1} \selectfont
-----\;\;\, 0.293 0.188 0.152 0.131 0.117 0.107 0.098 0.091 0.086 0.080 0.076 0.072 0.069 0.066 0.063 }\\
\hline
\end{tabular}
\caption{$\kappa$ as function of $M_{\chi_1^0}$ and $M_{\chi^\pm_1}$.}
\label{t:k}
\end{table}

\begin{table}[!h]
\center
\begin{tabular}{|c|c|}
\hline
$M_{\chi^\pm_1}$: &{\fontsize{10}{1} \selectfont
100.0 120.0 140.0 160.0 180.0 200.0 220.0 240.0 260.0 280.0 300.0 
 320.0 340.0  360.0 380.0 400.0 }\\
\hline \hline
$M_{\chi^0_1}$ & $M_{\chi^0_2}$  \\
\hline
  0 & {\fontsize{10}{1} \selectfont
105.9 125.1 144.4 163.7 183.2 202.7 222.3 241.9 261.5 281.2 300.9 320.6 340.3 360.1
 379.9 399.7 }\\
 10 & {\fontsize{10}{1} \selectfont
107.1 125.9 145.0 164.2 183.5 203.0 222.5 242.1 261.7 281.4 301.0 320.7 340.5 360.2 379.9 399.7 }\\
 20 & {\fontsize{10}{1} \selectfont
108.5 126.9 145.7 164.7 184.0 203.3 222.8 242.3 261.9 281.5 301.2 320.9 340.6 360.3 380.0 399.8 }\\
 30 & {\fontsize{10}{1} \selectfont
110.4 128.1 146.5 165.3 184.4 203.7 223.0 241.3 262.1 281.7 301.3 321.0 340.7 360.4 380.1 399.9 }\\
 40 & {\fontsize{10}{1} \selectfont
112.8 129.6 147.5 166.0 184.9 204.1 223.4 242.8 262.3 281.9 301.5 321.1 340.8 360.5 380.3 399.9 }\\
 50 & {\fontsize{10}{1} \selectfont
116.3 131.5 148.7 166.9 185.6 204.6 223.7 243.1 262.5 282.1 301.6 321.3 340.9 360.6 380.4 400.1 }\\
 60 & {\fontsize{10}{1} \selectfont
121.7 134.0 150.2 167.9 186.3 205.1 224.1 243.4 262.8 282.3 301.8 321.4 341.1 360.7 380.5 400.1 }\\
 70 & {\fontsize{10}{1} \selectfont
130.9 137.7 152.2 169.1 187.1 205.7 224.6 243.8 263.1 282.5 302.0 321.6 341.2 360.9 380.5 400.2 }\\
 80 & {\fontsize{10}{1} \selectfont
 \hspace{-8pt} -145.5 143.2 154.8 170.6 188.1 206.4 225.1 244.2 263.4 282.8 302.3 321.8 341.4 361.0 380.7 400.3 }\\
 90 & {\fontsize{10}{1} \selectfont
 \hspace{-5pt}-134.4 152.7 158.5 172.6 189.3 207.2 225.7 244.6 263.8 283.1 302.5 322.0 341.5 361.1 380.8 400.5 }\\
100 & {\fontsize{10}{1} \selectfont
-----\, -158.9 164.1 175.2 190.9 208.2 226.5 245.2 264.2 283.4 302.8 322.2 341.7 361.3 380.9 400.6 }\\
\hline
\end{tabular}
\caption{$M_{\chi^0_2}$ as function of $M_{\chi^0_1}$ and $M_{\chi^\pm_1}$.}
\label{t:m2}
\end{table}

\begin{table}[!h]
\center
\begin{tabular}{|c|c|}
\hline
$M_{\chi^\pm_1}$: &{\fontsize{10}{1} \selectfont
100.0 120.0 140.0 160.0 180.0 200.0 220.0 240.0 260.0 280.0 300.0 
 320.0 340.0  360.0 380.0 400.0 }\\
\hline \hline
$M_{\chi^0_1}$ & $-M_{\chi_3^0}$ \\
\hline
  0 & {\fontsize{10}{1} \selectfont
194.2 200.2 210.0 222.1 235.9 250.9 266.8 283.4 300.4 317.9 335.7 353.7 372.0 390.5 409.2 428.0 }\\
 10 & {\fontsize{10}{1} \selectfont
185.8 194.3 205.5 218.7 233.2 248.7 264.9 281.8 299.1 316.7 334.7 352.9 371.3 389.8 408.5 427.4 }\\
 20 & {\fontsize{10}{1} \selectfont
178.5 189.0 201.6 215.6 230.7 246.6 263.2 280.3 297.8 315.6 333.7 352.0 370.5 389.1 407.9 426.8 }\\
 30 & {\fontsize{10}{1} \selectfont
172.3 184.5 198.0 212.8 228.4 244.8 261.6 289.0 296.6 314.6 332.8 351.2 369.8 388.5 407.3 426.3 }\\
 40 & {\fontsize{10}{1} \selectfont
166.7 180.4 194.9 210.3 226.3 243.0 260.1 277.7 295.5 313.6 331.9 350.4 369.1 387.9 406.8 425.8 }\\
 50 & {\fontsize{10}{1} \selectfont
161.6 176.7 192.0 208.0 224.4 241.4 258.8 276.5 294.5 312.7 331.1 349.7 368.4 387.3 406.3 425.3 }\\
 60 & {\fontsize{10}{1} \selectfont
156.8 173.3 189.4 205.8 222.7 239.9 257.5 275.4 293.5 311.8 330.4 349.0 367.8 386.7 405.8 424.8 }\\
 70 & {\fontsize{10}{1} \selectfont
151.7 170.0 186.9 203.9 221.0 238.5 256.3 274.3 292.6 311.0 329.6 348.4 367.2 386.2 405.3 424.4 }\\
 80 & {\fontsize{10}{1} \selectfont
\hspace{-8pt} -150.7 166.8 184.6 202.0 219.5 237.2 255.2 273.4 291.7 310.3 329.0 347.7 366.7 385.7 404.8 423.9 }\\
 90 & {\fontsize{10}{1} \selectfont
\hspace{-5pt}-223.5 163.3 182.4 200.3 218.1 236.0 254.1 272.4 290.9 309.5 328.3 347.1 366.1 385.2 404.3 423.6 }\\
100 & {\fontsize{10}{1} \selectfont
-----\, -173.2 180.1 198.6 216.7 234.9 253.1 271.6 290.1 308.8 327.7 346.6 365.6 384.7 403.9 423.1 }\\\hline
\end{tabular}
\caption{$-M_{\chi_3^0}$ as function of $M_{\chi_1^0}$ and $M_{\chi^\pm_1}$. (The negative sign is chosen
only to keep the linewidths reasonably short.)}
\label{t:m3}
\end{table}

\begin{table}[h!]
\center
\begin{tabular}{|c|c|}
\hline
$M_{\chi^\pm_1}$: &{\fontsize{10}{1} \selectfont
100\; 120\; 140\; 160\; 180\; 200\; 220\; 240\; 260\; 280\; 300\; 
 320\; 340\;  360\; 380\; 400\;  }\\
\hline \hline
$M_{\chi^0_1}$ & $N_{2,3}$\\
\hline
  0 & {\fontsize{10}{1} \selectfont
.764 .751 .742 .735 .730 .726 .723 .720 .719 .717 .716 .714 .713 .713 .712 .711 }\\
 10 & {\fontsize{10}{1} \selectfont
.767 .753 .743 .736 .730 .726 .723 .721 .719 .717 .716 .715 .714 .713 .712 .711 }\\
 20 & {\fontsize{10}{1} \selectfont
.770 .754 .744 .736 .731 .727 .724 .721 .719 .717 .716 .715 .714 .713 .712 .711 }\\
 30 & {\fontsize{10}{1} \selectfont
.770 .755 .745 .737 .731 .727 .724 .720 .719 .717 .716 .715 .714 .713 .712 .711 }\\
 40 & {\fontsize{10}{1} \selectfont
.768 .755 .745 .738 .732 .727 .724 .721 .719 .718 .716 .715 .714 .713 .712 .711 }\\
 50 & {\fontsize{10}{1} \selectfont
.758 .753 .745 .738 .732 .728 .724 .722 .719 .718 .716 .715 .714 .713 .712 .712 }\\
 60 & {\fontsize{10}{1} \selectfont
.736 .747 .743 .737 .732 .728 .724 .722 .720 .718 .716 .715 .714 .713 .712 .712 }\\
 70 & {\fontsize{10}{1} \selectfont
.689 .734 .739 .736 .732 .728 .724 .722 .720 .718 .716 .715 .714 .713 .712 .712 }\\
 80 & {\fontsize{10}{1} \selectfont
.687 .707 .731 .733 .730 .727 .724 .722 .720 .718 .716 .715 .714 .713 .712 .712 }\\
 90 & {\fontsize{10}{1} \selectfont
.700 .653 .715 .727 .728 .726 .724 .722 .720 .718 .717 .715 .714 .713 .712 .712 }\\
100 & {\fontsize{10}{1} \selectfont
---- .690 .685 .717 .725 .725 .723 .721 .720 .718 .717 .715 .714 .713 .712 .712 }\\
\hline
\end{tabular}
\caption{$N_{2,3}$ as function of$M_{\chi_1^0}$ and $M_{\chi^\pm_1}$.}
\label{t:n23}
\end{table}

\begin{table}[h!]
\center
\begin{tabular}{|c|c|}
\hline
$M_{\chi^\pm_1}$: &{\fontsize{10}{1} \selectfont
\; 100\;\; 120\;\; 140\;\; 160\;\; 180\;\; 200\;\; 220\;\; 240\;\; 260\;\;\;
 280\;\; 300\;\; 320\;\;\; 340\;\;  360\;\; 380\;\; 400\;  }\\
\hline \hline\;
$M_{\chi^0_1}$ & $N_{2,4}$ \\
\hline
  0 & {\fontsize{10}{1} \selectfont
-.600 -.623 -.640 -.652 -.661 -.668 -.674 -.678 -.682 -.684 -.687 -.689 -.690 -.691 -.692 -.693 }\\
 10 & {\fontsize{10}{1} \selectfont
-.583 -.613 -.633 -.647 -.658 -.666 -.672 -.677 -.680 -.683 -.686 -.688 -.690 -.691 -.692 -.693 }\\
 20 & {\fontsize{10}{1} \selectfont
-.562 -.600 -.625 -.642 -.654 -.663 -.670 -.675 -.679 -.682 -.685 -.687 -.689 -.691 -.692 -.693 }\\
 30 & {\fontsize{10}{1} \selectfont
-.535 -.585 -.615 -.636 -.650 -.660 -.668 -.682 -.678 -.681 -.684 -.687 -.688 -.690 -.691 -.692 }\\
 40 & {\fontsize{10}{1} \selectfont
-.499 -.565 -.604 -.628 -.645 -.656 -.665 -.671 -.676 -.680 -.683 -.686 -.688 -.689 -.691 -.692 }\\
 50 & {\fontsize{10}{1} \selectfont
-.450 -.540 -.589 -.619 -.639 -.652 -.662 -.669 -.675 -.679 -.682 -.685 -.687 -.689 -.690 -.692 }\\
 60 & {\fontsize{10}{1} \selectfont
-.380 -.506 -.571 -.608 -.632 -.647 -.659 -.667 -.673 -.677 -.681 -.684 -.686 -.688 -.690 -.691 }\\
 70 & {\fontsize{10}{1} \selectfont
 .279 -.459 -.546 -.594 -.623 -.642 -.655 -.664 -.671 -.676 -.680 -.683 -.686 -.688 -.689 -.691 }\\
 80 & {\fontsize{10}{1} \selectfont
 .616 -.391 -.513 -.576 -.612 -.635 -.650 -.661 -.668 -.674 -.678 -.682 -.685 -.687 -.689 -.690 }\\
 90 & {\fontsize{10}{1} \selectfont
 .648 -.294 -.468 -.553 -.599 -.627 -.645 -.657 -.665 -.672 -.677 -.681 -.684 -.686 -.688 -.690 }\\
100 & {\fontsize{10}{1} \selectfont
-----  .640 -.402 -.521 -.582 -.616 -.638 -.652 -.662 -.670 -.675 -.679 -.683 -.685 -.687 -.689 }\\\hline
\end{tabular}
\caption{$N_{2,4}$ as function of$M_{\chi_1^0}$ and $M_{\chi^\pm_1}$.}
\label{t:n24}
\end{table}

\begin{table}
\center
\begin{tabular}{|c|c|}
\hline
$M_{\chi^\pm_1}$: &{\fontsize{10}{1} \selectfont
100\; 120\; 140\; 160\; 180\; 200\; 220\; 240\; 260\; 280\; 300\; 
 320\; 340\;  360\; 380\; 400\;  }\\
\hline \hline
$M_{\chi^0_1}$ & $N_{3,3}$ \\
\hline
  0 & {\fontsize{10}{1} \selectfont
.571 .600 .621 .636 .648 .658 .665 .671 .675 .679 .683 .685 .687 .689 .691 .693 }\\
 10 & {\fontsize{10}{1} \selectfont
.591 .614 .632 .645 .655 .662 .669 .674 .678 .681 .684 .687 .689 .690 .692 .693 }\\
 20 & {\fontsize{10}{1} \selectfont
.609 .627 .641 .652 .660 .667 .672 .677 .680 .683 .686 .688 .690 .691 .693 .694 }\\
 30 & {\fontsize{10}{1} \selectfont
.625 .638 .649 .658 .665 .671 .675 .659 .682 .685 .687 .689 .691 .692 .694 .695 }\\
 40 & {\fontsize{10}{1} \selectfont
.638 .648 .656 .663 .669 .674 .678 .681 .684 .686 .689 .690 .692 .693 .694 .695 }\\
 50 & {\fontsize{10}{1} \selectfont
.650 .656 .662 .668 .673 .677 .680 .683 .686 .688 .690 .691 .693 .694 .695 .696 }\\
 60 & {\fontsize{10}{1} \selectfont
.661 .663 .668 .672 .676 .680 .683 .685 .687 .689 .691 .692 .693 .695 .696 .696 }\\
 70 & {\fontsize{10}{1} \selectfont
.672 .670 .673 .676 .679 .682 .685 .687 .689 .690 .692 .693 .694 .695 .696 .697 }\\
 80 & {\fontsize{10}{1} \selectfont
.583 .676 .677 .679 .682 .684 .686 .688 .690 .691 .693 .694 .695 .696 .697 .697 }\\
 90 & {\fontsize{10}{1} \selectfont
.374 .683 .681 .682 .684 .686 .688 .690 .691 .692 .694 .695 .696 .696 .697 .698 }\\
100 & {\fontsize{10}{1} \selectfont
---- .548 .685 .685 .686 .688 .689 .691 .692 .693 .694 .695 .696 .697 .698 .698 }\\
\hline
\end{tabular}
\caption{$N_{3,3}$ as function of $M_{\chi^0_1}$ and $M_{\chi^\pm_1}$.}
\label{t:n33}
\end{table}

\begin{table}[h!]
\center
\begin{tabular}{|c|c|}
\hline
$M_{\chi^\pm_1}$: &{\fontsize{10}{1} \selectfont
100\; 120\; 140\; 160\; 180\; 200\; 220\; 240\; 260\; 280\; 300\; 
 320\; 340\;  360\; 380\; 400\; }\\
\hline \hline
$M_{\chi^0_1}$ & $N_{3,4}$ \\
\hline
  0 & {\fontsize{10}{1} \selectfont
.468 .515 .550 .578 .600 .616 .630 .641 .649 .657 .662 .667 .672 .675 .678 .681 }\\
 10 & {\fontsize{10}{1} \selectfont
.492 .533 .564 .589 .608 .623 .635 .645 .653 .659 .665 .669 .673 .677 .680 .682 }\\
 20 & {\fontsize{10}{1} \selectfont
.513 .549 .576 .598 .615 .629 .640 .648 .656 .662 .667 .671 .675 .678 .681 .683 }\\
 30 & {\fontsize{10}{1} \selectfont
.532 .563 .587 .606 .622 .634 .644 .625 .659 .664 .669 .673 .676 .679 .682 .684 }\\
 40 & {\fontsize{10}{1} \selectfont
.549 .575 .597 .614 .627 .639 .648 .655 .661 .666 .671 .675 .678 .681 .683 .685 }\\
 50 & {\fontsize{10}{1} \selectfont
.565 .587 .605 .620 .633 .643 .651 .658 .664 .668 .673 .676 .679 .682 .684 .686 }\\
 60 & {\fontsize{10}{1} \selectfont
.581 .597 .613 .626 .638 .647 .654 .661 .666 .670 .674 .677 .680 .683 .685 .687 }\\
 70 & {\fontsize{10}{1} \selectfont
.597 .607 .620 .632 .642 .650 .657 .663 .668 .672 .676 .679 .681 .684 .686 .687 }\\
 80 & {\fontsize{10}{1} \selectfont
-.137 .617 .627 .637 .646 .654 .660 .665 .670 .674 .677 .680 .682 .685 .686 .688 }\\
 90 & {\fontsize{10}{1} \selectfont
.252 .627 .633 .642 .650 .657 .662 .667 .672 .675 .678 .681 .683 .685 .687 .689 }\\
100 & {\fontsize{10}{1} \selectfont
---- .155 .640 .646 .653 .659 .665 .669 .673 .677 .680 .682 .684 .686 .688 .689 }\\
\hline
\end{tabular}
\caption{$N_{3,4}$ as function of$M_{\chi^0_1}$ and $M_{\chi^\pm_1}$.}
\label{t:n34}
\end{table}

\end{document}